\documentclass[conference]{IEEEtran}

\IEEEoverridecommandlockouts
\usepackage{amsmath,amssymb,amsfonts}
\usepackage{algorithmic}
\usepackage{graphicx}
\usepackage{textcomp}
\usepackage[dvipsnames]{xcolor}

\usepackage{algorithm}
\usepackage{array}
\usepackage{stfloats}
\usepackage{url}
\usepackage[utf8]{inputenc} % spectial character
\usepackage{bm} % bold italic
\usepackage{caption}
\usepackage{subcaption}
\usepackage{booktabs} % table formating
\usepackage{xifthen} % subequation continous numbering
\usepackage{cite}
\usepackage{hyperref}
\newcommand{\ut}[1]{\bm{u}_{#1}}
\newcommand{\utp}[1]{\bm{u}_{#1}^\prime}
\newcommand{\utt}[1]{\tilde{\bm{u}}_{#1}}
\newcommand{\vt}[1]{\bm{v}_{#1}}
\newcommand{\vtp}[1]{\bm{v}_{#1}^\prime}
\newcommand{\xt}[1]{\bm{x}_{#1}}
\newcommand{\xtp}[1]{\bm{x}_{#1}^\prime}
\newcommand{\mut}[1]{\bm{\mu}_{#1}}
\newcommand{\uth}[1]{\mathbf{u}_{#1}}
\newcommand{\vth}[1]{\mathbf{v}_{#1}}

\newcommand{\norm}[1]{\lVert#1\rVert}
\newcommand{\ceil}[1]{\lceil#1\rceil}
\newcommand{\Ga}[0]{\mathbf{G}_{457}}

\newcommand{\Gb}[0]{\mathbf{G}_{537}}
\newcommand{\Gbf}[0]{\mathbf{G}_{357}}

\newcommand{\ebp}[0]{\epsilon_{\textrm{BP}}}
\newcommand{\ebpf}[0]{\epsilon_{\textrm{BP,full}}}
\newcommand{\ebpw}[0]{\epsilon_{\textrm{BP,wd}}}
\newcommand{\emap}[0]{\epsilon_{\textrm{MAP}}}
\newcommand{\eh}[0]{\epsilon_{\textrm{h}}}
\newcommand{\et}[0]{\epsilon_{\textrm{t}}}

\newcommand{\ev}[0]{\varepsilon_{\textrm{v}}}

\newcommand{\head}[1]{^{\left(#1\right)}} 
\newcommand{\vhead}[1]{^{\left<#1\right>}} 
\newcommand{\U}[0]{\textrm{U}}
\renewcommand{\L}[0]{\textrm{L}}

\newcommand{\fu}[1]{ f_{#1}^{\U} }
\newcommand{\fl}[1]{ f_{#1}^{\L} }
\newcommand{\fphi}[1]{ f^{\phi}_{#1} }

\newcounter{example} %{\medskip \rmfamily}
\newcounter{remark} \newenvironment{remark}[1][]{\refstepcounter{remark}\par\textbf{Remark~\theremark: #1}}{~ \hfill $\blacksquare$ }
\newcounter{define} 
\begin{document}
	\title{Half Spatially Coupled Turbo-Like Codes
	}
	\author{ 
		\IEEEauthorblockN{Xiaowei Wu\IEEEauthorrefmark{1}, Lei Yang\IEEEauthorrefmark{1}, Min Qiu\IEEEauthorrefmark{2}, Chong Han\IEEEauthorrefmark{3}, Jinhong Yuan\IEEEauthorrefmark{2}}
		\IEEEauthorblockA{\IEEEauthorrefmark{1}Technology and Engineering Center for Space Utilization, Chinese Academy of Sciences, Beijing, China. \\Email: \{wuxiaowei, yang.lei\}@csu.ac.cn}		
		\IEEEauthorblockA{\IEEEauthorrefmark{2}University of New South Wales, Sydney, Australia. Email: \{min.qiu, j.yuan\}@unsw.edu.au} 	
		\IEEEauthorblockA{\IEEEauthorrefmark{3}Shanghai Jiao Tong University, Shanghai, China. Email: \{chong.han\}@sjtu.edu.cn} 
	}
	\maketitle
	\begin{abstract}
		This paper presents a new class of spatially coupled turbo-like codes (SC-TCs), namely half spatially coupled braided convolutional codes (HSC-BCCs) and half spatially coupled parallel concatenated codes (HSC-PCCs). Different from the conventional SC-TCs, the proposed codes have simpler and deterministic coupling structures. Most notably, the coupling of HSC-BCCs is performed by re-encoding the whole coupling sequence in the component encoder of one time instant, rather than spreading the coupling bits to component encoders of multiple time instants. This simplification not only addresses the window decoding threshold loss issue in existing BCCs, but also allows the proposed codes to attain very close-to-capacity performance with a coupling memory as small as 2. Both theoretical and numerical results are provided to demonstrate the performance advantages of the proposed codes over existing spatially coupled codes.
	\end{abstract}
	\begin{IEEEkeywords}
		Spatial coupling, Turbo-like codes, Braided convolutional codes, Sliding window decoding 
	\end{IEEEkeywords}
	
	\section{Introduction} 
	Spatial coupling, initially introduced in \cite{Felstrom-convLDPC} to design convolutional low-density parity check (LDPC) codes, is a technique to construct capacity-achieving codes and has attracted extensive investigations in the last decade \cite{Mitchell-scldpc,sc-ldpc_bicmid,Xie_scldpc}.
	Spatial coupling is also applied to concatenated convolutional codes, resulting in various spatially coupled turbo-like codes (SC-TCs) \cite{Zhang-bcc,Moloudi-scTurbo,Min-picppcTC,min_gscpcc,Yang-hybrid_scc_2,Yang-picTurbo,Wu-picdTC}.
	In \cite{min_gscpcc}, generalized spatially coupled parallel concatenated codes (GSC-PCCs) were proved to hold the threshold saturation effect and capacity-achieving. As various SC-TCs have already been shown to have close-to-capacity decoding thresholds \cite{Moloudi-scTurbo,Min-picppcTC,min_gscpcc,Yang-hybrid_scc_2}, the research motivations in recent works are not only limited to designing code ensembles with optimized decoding thresholds, but also towards performance-complexity trade-offs and practical implementations  \cite{Graell-bcc_errfloor,Mahd-scscc_hard,Zhu-bcsoc,Farooq-scscc,Wang-hybrid_scc_errorfloor}. 
	
	Spatially coupled braided convolutional codes (SC-BCCs), first proposed in \cite{Zhang-bcc}, are constructed by using the component multi-binary recursive systematic convolutional (RSC) codes of one time instant to re-encode the parity bits from the preceding time instants. Various works have been carried out on performance analysis \cite{Moloudi-bcc_de_bec,Farooq-bcc_de_awgn,Moloudi-bcc_weight}, decoding scheme \cite{zhu-bcc_window,Zhu-bcc_dec}, and code design extensions  \cite{Moloudi-scTurbo, Zhu-bcsoc}. Notably, \cite{Moloudi-scTurbo} presents type-2 BCCs as an extension of the original design (type-1 BCCs) in \cite{Zhang-bcc}. This new subclass is proven to exhibit threshold saturation, and its belief propagation (BP) decoding threshold is close to capacity over binary erasure channels (BEC).
	
	Although having many attractive features, type-1 BCCs suffer from decoding threshold degradation under window decoding, even with a sufficiently large window size \cite{zhu-bcc_window}. Type-2 BCCs \cite{Moloudi-scTurbo} require a relatively large coupling memory to resolve the window decoding threshold loss problem. The randomized coupling structure also makes type-2 BCCs difficult to implement, and does not always guarantee good finite-length performance. Some existing schemes, such as partially information coupled turbo codes (PIC-TCs) \cite{Yang-picTurbo,Wu-picdTC} and GSC-PCCs \cite{min_gscpcc}, need a large coupling memory to approach capacity. This motivates us to investigate designs that can achieve close-to-capacity performance under window decoding while having a simplified and deterministic coupling structure.
	
	In this paper, we introduce the class of half spatially coupled turbo-like codes. We start with the simplification of SC-PCC ensembles and provide a different representation on the code structures to facilitate more effective coupling designs. The proposed half SC-PCCs (HSC-PCCs) retain the same decoding threshold while having a simple and deterministic coupling structure. Building upon the half coupling structures, we then construct half SC-BCCs (HSC-BCCs). Specifically, the source information bits and parity bits of one time instant are each re-encoded by RSC codes of successive but different time instants to avoid high dependency between factor nodes caused by short decoding cycles. Despite the simple coupling structures, HSC-BCCs outperform type-1 and type-2 BCCs in terms of decoding thresholds and finite-length error performance.
	
	\textbf{Notations and Definitions: }
	Scalars and vectors are written in light and bold letters, respectively, e.g. $x$ and $\xt{}$. $\norm{\xt{}}$ gives the number of elements in a vector $\xt{}$. 
	For an RSC encoder with $k\geq 1$ inputs, we use $\uth{}\head{i}$ to denote the $i$-th input. For an SC-TC with rate-$\frac{k}{k+1}$ component RSCs and a coupling length $T$, we use $\uth{t}\head{i,\U}$ and $\uth{t}\head{i,\L}$ to represent the $i$-th input of the upper and lower RSC encoders at time $t\in\{1,\ldots,T\}$, where $i\in\{1,\ldots,k\}$.
	We write $\uth{t}\head{i,\U} = \boldsymbol{x}$ if a sequence $\boldsymbol{x}$ is encoded by the upper RSC encoder at its $i$-th input at time $t$. $\Pi(.)$ denote an interleaving function.
	
	\section{Preliminaries} 
	In this section, we review type-1 and type-2 BCCs. We use rate-1/3 BCCs built with rate-$2/3$ component RSCs, as in \cite{Moloudi-scTurbo}. We also presume that interleaves are time-invariant.
	
\begin{figure*}[t]
		\centering
		\subcaptionbox{ \label{fig:bcc1} }  	{\includegraphics[width=5.1cm]{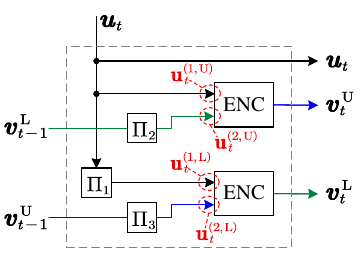} \vspace{10pt}}
%		 \hspace{-12pt}
		\subcaptionbox{ \label{fig:bcc2} } 		{\includegraphics[width=6.3cm]{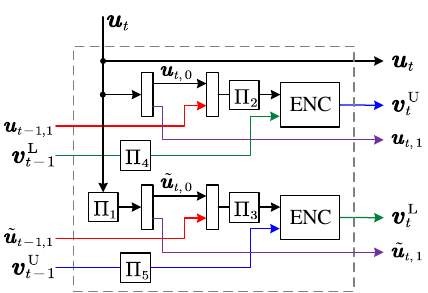}} 
%		\hspace{-12pt}
		\subcaptionbox{ \label{fig:scpcc} } 	{\includegraphics[width=5.9cm]{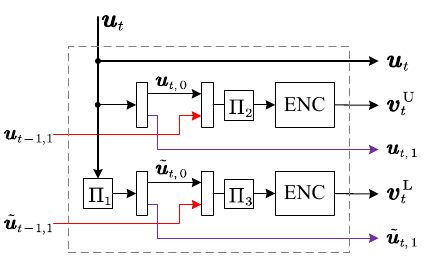} \vspace{10pt}}
		\caption{Encoders of (a) a type-1 BCC, (b) a type-2 BCC, and (c) an SC-PCC with $m = 1$ at time $t$.}
		\label{fig:SC_TC}
	\end{figure*}

	Type-1 BCCs \cite{Zhang-bcc} are constructed by using parallel RSC codes at time $t$ to encode the source information sequence $\ut{t}$ together with the parity bits from previous time instants. The type-1 BCC encoder with coupling memory ${m=1}$ is depicted in Fig. \ref{fig:bcc1}, where the input sequences to the upper and lower encoders at time $t$ are $[ \uth{t}\head{1,\U}=\ut{t}, \uth{t}\head{2,\U}=\Pi_2(\vt{t-1}^\L) ]$, and $[ \uth{t}\head{1,\L}=\Pi_1(\ut{t}),\uth{t}\head{2,\L}=\Pi_3(\vt{t-1}^\U) ]$, respectively. As shown in \cite{zhu-bcc_window}, type-1 BCCs suffer from a noticeable decoding threshold degradation under window decoding compared to full decoding, even under a sufficiently large window size $w$.
	
	Type-2 BCCs \cite{Moloudi-scTurbo} are constructed by combining the coupling structure of type-1 BCCs and SC-PCCs so that both information and parity bits are coupled. The encoder of type-2 BCCs with coupling memory $m=1$ is depicted in Fig. \ref{fig:bcc2}. Type-2 BCCs show better full decoding threshold than type-1 BCCs, but still suffer from window decoding threshold loss unless under a large coupling memory, e.g., $m\geq5$.
	
	\begin{remark}
		Aware that decreasing the window size $w$ leads to decoding threshold degradation, the window decoding threshold $\ebpw$ discussed in this paper refers to the threshold achieved with a sufficiently large $w$ such that the threshold improvement given by further increasing $w$ is negligible. We use ${w=10m}$ in the DE computations for convenience. Of course, such a large size is not necessary in practice.
	\end{remark}
	
	\section{Half Spatially Coupled Turbo-Like Codes}  
	In this section, we introduce HSC-PCCs and HSC-BCCs, which simplify the coupling structures of the SC-PCCs and type-2 BCCs in \cite{Moloudi-scTurbo}. The benefits of the proposed half coupling will be revealed in Sec. \ref{sec:analysis}.
	
	\subsection{A Different View on SC-PCCs}
	We start with the conventional SC-PCCs and provide a different view of their coupling structure, which allows us to derive the proposed coding structure.
	The encoder of SC-PCC${(m=1)}$ is shown in Fig. \ref{fig:scpcc}. At time $t$, $\ut{t}$ is segmented into $m+1$ equal-length subsequences $\ut{t}= [\ut{t,0},\ut{t,1},\ldots,\ut{t,m}]$. Similarly, the permutation of $\ut{t}$ is decomposed into $m+1$ equal-length subsequences $\utt{t} = \Pi_1(\ut{t}) = [ \utt{t,0},\ldots, \utt{t,m}]$. Then, the upper and lower RSC encoders input
	\begin{align}
		\left\lbrace \begin{aligned}
			&\uth{t}\head{1,\U} =  \Pi_2\big(\left[\ut{t-m,m},\hdots,\ut{t,0}\right]\big)  \\
			& \uth{t}\head{1,\L} = \Pi_3\big(\left[ \utt{t-m,m},\hdots, \utt{t,0}\right]\big) \\
		\end{aligned}
		\label{eq:scpcc0}
		\right. .
	\end{align}
	Note that $\utt{t,i}$ is not equivalent to the reordered copy of $\ut{t,i}$ for $i=0,\hdots,m$.
	
\begin{figure*}[t]
		\centering
		\subcaptionbox{ \label{fig:factor_graph_scpcc1} }	{\includegraphics[height=4.5cm]{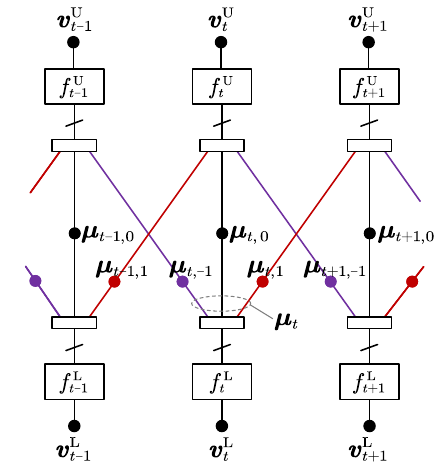}} \hspace{5pt}
		\subcaptionbox{ \label{fig:factor_graph_hscpcc1} } 	{\includegraphics[height=4.5cm]{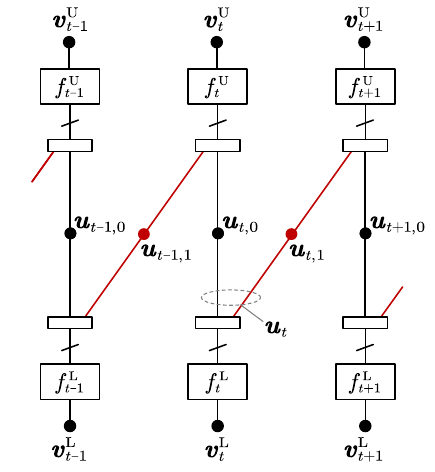}} \hspace{5pt}
		\subcaptionbox{ \label{fig:HSCPCC_encoder_m} } {\includegraphics[width=6cm]{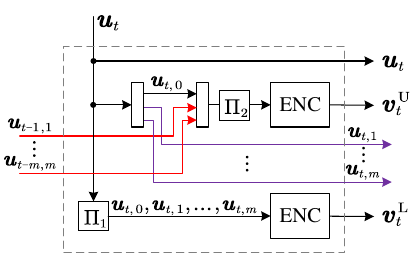}\vspace{20pt}}   
		\caption{Compact graphs of (a) SC-PCC$(m=1)$ with the equivalent expressions, and (b) single-sided SC-PCC$(m=1)$. (c) Encoder of single-sided SC-PCC$(m\geq 1)$.}
		\label{fig:factor_graph_scpcc}
	\end{figure*}

	Let $\ut{t,(a,b)}$ represent information bits that appeared in both $\ut{t,a}$ and $\utt{t,b}$, where $a,b\in\{0,\hdots,m\}$. The bits that are encoded as a part of $\uth{t+i}\head{1,\U}$ and $\uth{t}\head{1,\L}$ are written together as $\mut{t,i}=[\ut{t^\prime,(a,b)}]^t_{t'=t-m}$, where $a-b = i\in\{0,\hdots,m\}$.	With this representation, \eqref{eq:scpcc0} can be rewritten as
	\begin{align}
		\left\lbrace \begin{aligned}
			&\uth{t}\head{1,\U} = \Pi \big( [\mut{t-i,i}]^m_{i=-m} \big)   \\
			&\uth{t}\head{1,\L} = [\mut{t,i}]^m_{i=-m} = \mut{t} \\
		\end{aligned}
		\right. ,
		\label{eq:scpcc_eq}
	\end{align}
	where $\mut{t}$ is referred to as the source information of time $t$ in this equivalent representation, $\mut{t,0}$ is the uncoupled subsequence of $\mut{t}$, and $\mut{t,i\neq0}$ is the coupled subsequence being encoded at the upper encoder of time $t+i$ and the lower encoder of time $t$. As an example, the compact graph of SC-PCC$(m=1)$ is redrawn as Fig. \ref{fig:factor_graph_scpcc1}, where $\mut{t,-1}= \ut{t-1,(0,1)}$, $\mut{t,0}= [\ut{t-1,(1,1)},\ut{t,(0,0)}]$, and $\mut{t,1}= \ut{t,(1,0)}$. 
	
	In the original SC-PCCs encoder from \cite{Moloudi-scTurbo}, the ratio of $\mut{t,i}$ over $\mut{t}$, expressed as $\lambda_i=\frac{\norm{\mut{t,i}}}{\norm{\mut{t}}}$, is completely random at finite blocklengths. Consequently, $\lambda_i$ could be very small, such that the decoding of subblock $i$ would not leverage the coupling gain. With infinite blocklength and an ideal random interleaver, the average ratio of $\mut{t,i}$ becomes ${\bar{\lambda}_i = \frac{m-\text{abs}(i)+1}{(m+1)^2}}$, where $\bar{\lambda}_i > \bar{\lambda}_j$ for any $\text{abs}(i)<\text{abs}(j)$ and $i,j\in \{0,\hdots,m\}$. The average coupling ratio is ${\bar{\lambda}=1-\frac{\norm{\mut{t,0}}}{\norm{\mut{t}}}=\frac{m}{m+1}}$. 
	With the equivalent representation \eqref{eq:scpcc_eq}, $\lambda_i$ can be made deterministic even at finite blocklengths. This allows us to obtain new designs by varying the coupling ratio.
	
	\begin{figure}[t]
		\centering
		\subcaptionbox{\label{fig:factor_graph_hscpcc2}} {\includegraphics[width=6cm]{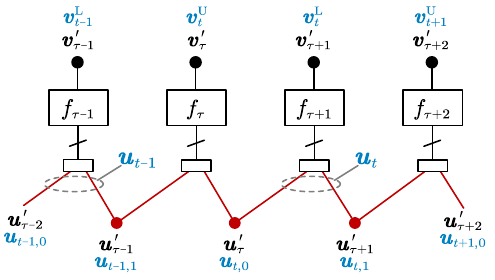}}
		\subcaptionbox{\label{fig:HSCPCC_encoder_1}} {\includegraphics[width=8.5cm]{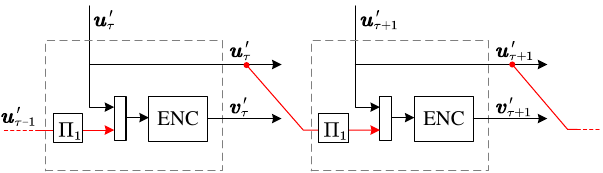} } 
		\caption{HSC-PCC$(m=1)$ in the folded form: (a) compact graph, and (b) encoder. }
		\label{fig:hscpcc}
	\end{figure}
	
	\subsection{Half SC-PCCs}
	By forcing ${\lambda_i=0}$ for ${i=-m,\hdots,-1}$, we get  a subclass of SC-PCCs that only pass  the coupled subsequences in the forward direction, i.e.
	\begin{align}
		\left\lbrace \begin{aligned}
			&\uth{t}\head{1,\U}= \Pi \big([\ut{t,0},\ut{t-1,1},\hdots,\ut{t-m,m}]) \\
			&\uth{t}\head{1,\L}=\ut{t}=[\ut{t,0},\ut{t,1},\hdots,\ut{t,m}] \\
		\end{aligned}
		\right.
	\end{align}
	This new subclass is referred to as the single-sided SC-PCCs. 
	Fig. \ref{fig:factor_graph_hscpcc1} depicts the compact graph for $m=1$, and Fig. \ref{fig:HSCPCC_encoder_m} shows the encoder for $m \geq 1$. 
	In Sec. \ref{sec:de_hscpcc}, we will show that the single-sided SC-PCCs with arbitrary coupling memory $m \geq 1$ exhibit  threshold saturation when $\ut{t}$ is  evenly split such that $\lambda=\frac{m}{m+1}$ and $\lambda_i=\frac{1}{m+1}$ for $i=0,1,\hdots,m$.
	
	Consider the single-sided SC-PCCs with $m=1$ and $\lambda_0=\lambda_1=0.5$. In this particular case, each component RSC code shares half of the information inputs with its neighboring component RSC code, so we refer to it as \emph{half} SC-PCCs (HSC-PCCs).
	We can illustrate the compact graph of HSC-PCC$(m=1)$ in a folded form, as depicted in Fig. \ref{fig:factor_graph_hscpcc2}, and the corresponding encoder implementation is shown in Fig. \ref{fig:HSCPCC_encoder_1}. The encoding process is as follows:
	\begin{enumerate}
		\item Given a total blocklength of $KT$, the source information is divided into $2T$ sequences of length $K/2$, denoted by $\utp{1}, \utp{2}, \hdots, \utp{2T}$, where the last sequence $\ut{2T}$ is force to be all-zero for termination.
		\item For $\tau=1,2,\hdots,2T$, the $\tau$-th RSC encoder takes
		\begin{align}
			\uth{\tau}\head{1}=[\utp{\tau},\Pi_1(\utp{\tau-1})]
		\end{align}
		as inputs, and outputs the parity sequence $\vtp{\tau}$, resulting the codeword $\xtp{\tau}=[\utp{\tau}, \vtp{\tau}]$.
		\item For $\tau=1,2,\hdots,2T-1$, $\xtp{\tau}$ is transmitted. For $\tau=2T$, only $\vtp{\tau}$ is transmitted.
	\end{enumerate}	
	The index $\tau$ refers to a half time instant: each information sequence $\utp{\tau}$ corresponds to either $\ut{t,0}$ or $\ut{t,1}$ of single-sided SC-PCC$(m=1)$; each parity sequence $\vtp{\tau}$ corresponds to either $\vt{t}^\U$ or $\vt{t}^\L$ of single-sided SC-PCC$(m=1)$. Thus, the code rate of HSC-PCC$(m=1)$ is  $R=\frac{T-0.5}{3T-0.5}\overset{\scalebox{0.6}{$T\!\rightarrow\!\infty$}}{=}1/3$.

	\begin{figure}[t]
		\centering
		\subcaptionbox{Turbo form \label{fig:factor_graph_hbcc1}}  {\includegraphics[height=4cm]{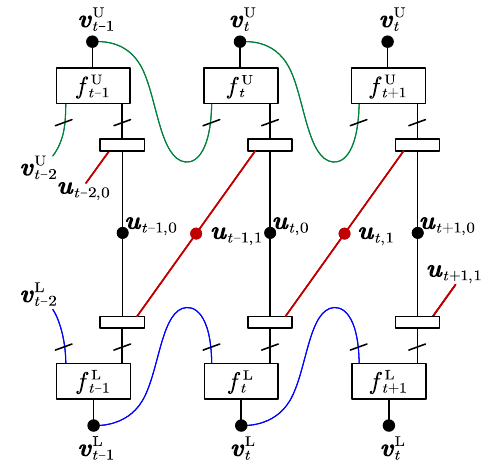} }
		\subcaptionbox{Folded form \label{fig:factor_graph_hbcc_m1s2_2} }  {\includegraphics[height=3cm]{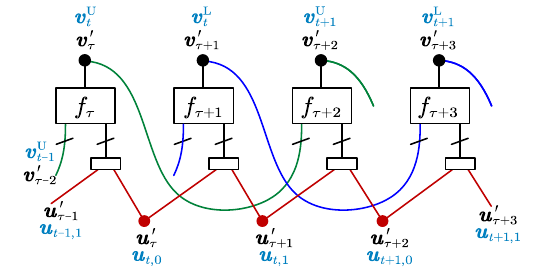} }
		\caption{Compact graphs of proposed HSC-BCC$(\delta =2)$.}
		\label{fig:factor_graph_hbcc}
		%	\end{figure}
	%	
	%	\begin{figure}[t]
		%		\centering
		\vspace{10pt}
		\subcaptionbox{$\delta = 2$}{\includegraphics[width=\linewidth]{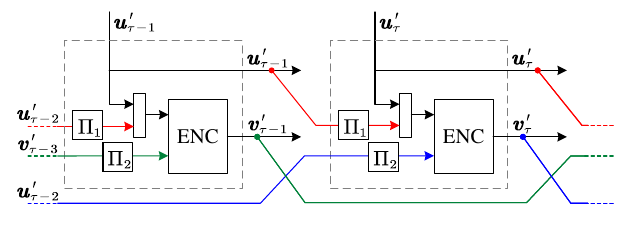}}
		\subcaptionbox{$\delta \geq 2$}{\includegraphics[width=5cm]{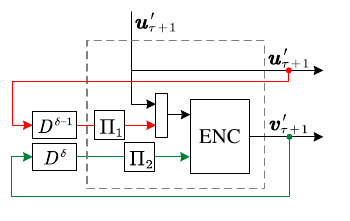}}
		\caption{Encoder block diagram of HSC-BCCs. } \label{fig:HSC-BCC_encoder_1}
	\end{figure}
	
	\subsection{Half SC-BCCs}
	HSC-BCCs are constructed by replacing the component rate-1/2 RSC code of HSC-PCC$(m=1)$ with the rate-2/3 duo-binary RSC code and using the additional input to re-encode the parity sequence from the previous time instant. 
	At time $\tau$, the component RSC encoder takes
	\begin{align}
		\left\lbrace \begin{aligned}
			&\uth{\tau}\head{1}=[ \utp{\tau}, \Pi_1 ( \utp{\tau-\delta +1}) ]  \\
			&\uth{\tau}\head{2}=\Pi_2( \vtp{\tau-\delta} )
		\end{aligned}
		\right. ,
	\end{align}
	where $\delta \geq 2$ is referred to as the delay factor of HSC-BCCs. The coupling memory of HSC-BCC$(\delta)$ in the turbo form is $m=\ceil{\delta /2}$. The coupling is terminated by setting the last $\delta$ information blocks $\utp{2T-\delta+1},\hdots,\utp{2T}$ to zero, so that the code rate is $ R=\frac{T-0.5\delta}{3T-0.5\delta} \overset{\scalebox{0.6}{$T\!\rightarrow\!\infty$}}{=}1/3$.

\begin{figure}[t]
		\centering
		\subcaptionbox{ Type-2 BCC \label{fig:factor_graph_bcc2} } [0.44\columnwidth] {\includegraphics[height=4cm]{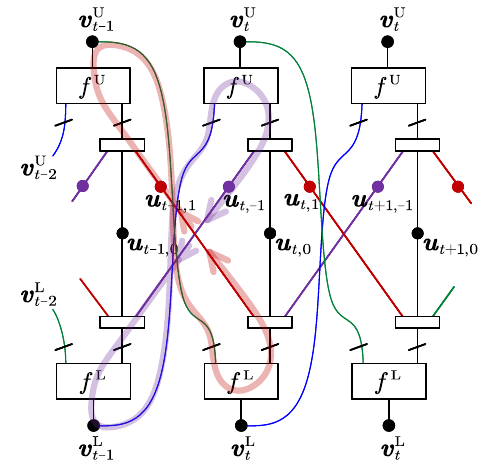}}
		\hspace{10pt}
		\subcaptionbox{ An SC-BCC variant \label{fig:factor_graph_hbcc1_notgood}} [0.44\columnwidth]  {\includegraphics[height=4cm]{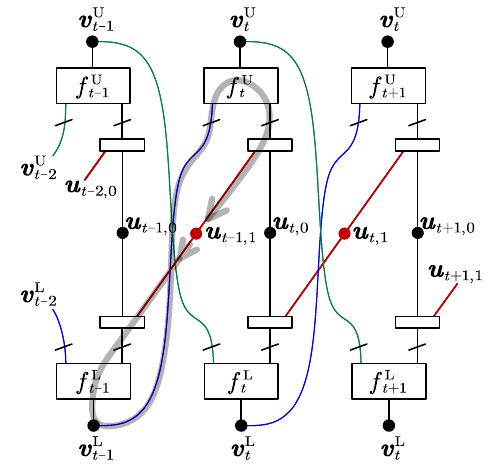}} 
		\caption{Decoding cycles on SC-BCC$(m=1)$.}
		\label{fig:factor_graph_hbcc2}
	\end{figure}

	The proposed HSC-BCCs do not re-encode $\ut{\tau}$ and $\vt{\tau}$ at the same time instants $\tau+\delta$ to avoid high dependency between the decoding messages given by the $\tau$-th and $(\tau+\delta)$-th component RSC code. 	
	Taking ${m=1}$ as an example, Fig. \ref{fig:factor_graph_hbcc} depicts HSC-BCC${(\delta=2)}$, and Fig. \ref{fig:factor_graph_hbcc2} depicts an SC-BCC variant that directly combines the coupling structure of type-1 BCC${(m=1)}$ and HSC-PCC${(m=1)}$. One can clearly see that such a variant forms a very short cycle between every two adjacent time instants. To be specific, both $\ut{t-1,1}$ and $\vt{t-1}^\L$ are connected between $\fl{t-1}$ and $\fu{t}$, so that decoding messages will propagate along the cycle $\ut{t-1,1} \rightarrow \fl{t-1} \rightarrow \vt{t-1}^\L \rightarrow \fu{t} \rightarrow \ut{t-1,1}$. Thus, the decoding outcomes of $\fu{t}$ have a much higher dependency on $\fl{t-1}$ than on $\fl{t}$. 
	Likewise, type-2 BCC$(m=1)$ also has two short cycles between every two adjacent time instants so that $\fu{t}$ has high dependency on $\fl{t-1}$ and $\fl{t+1}$ but very low dependency $\fl{t}$. HSC-BCC$(\delta=2)$ avoids such types of short cycles by respectively re-encoding the parity sequences by the component RSC encoder on the same side, instead of on the opposite side. This also improves the regularity of the coupling structure in the folded compact graph, allowing low complexity implementations of the encoder and decoder. 
	
	\begin{remark}
		As shown in Fig. \ref{fig:HSC-BCC_encoder_1}, the encoder of HSC-BCC$(\delta \geq 2)$ is much simpler than type-2 BCC${(m=1)}$ in Fig. \ref{fig:bcc2}. Moreover, unlike many existing SC-TCs that split each coupled sequence into $m$ (or $m+1$) segments to attain a coupling memory of $m$, HSC-BCCs increase the delay factor $\delta$ without any additional segmentation operation. This reduces the (de-)multiplexing operations during encoding. Correspondingly, the complexity of the decoder implementation is also greatly reduced.
	\end{remark}
	
	\section{Performance Analysis}\label{sec:analysis} 
	In this section, density evolution over BEC is used to illustrate the window decoding threshold loss of the existing BCCs. Then, the window decoding thresholds of HSC-PCCs and HSC-BCCs are presented. 
	Let $\varepsilon$ be the BEC erasure probability. We consider that parity bits are punctured for attain rates above 1/3, and $\ev$ denotes the erasure probability of the received parity bits with puncturing being taken into account. The input-output relations at the factor node $f$ can be described by a set of transfer functions $\{F_1(\cdot),F_2(\cdot),F_3(\cdot)\}$, and the relevant notations are given in Table \ref{tab:ND}. 	
	
	\begin{table}[h]
		\centering
		\caption{Notations and definitions for DE}
		\label{tab:ND}
		\begin{tabular}{ll}
			\toprule
			Notation & Definition \\		\midrule
			$F_{1}$$\big/$$F_{2}$$\big/$$F_{3}$ & Transfer function of $f$ for $\uth{}\head{1}$$\big/$$\uth{}\head{2}$$\big/$$\vth{}$ \\
			$\bar{\mathsf{p}}_{i ,\phi,t}\vhead{\ell}$$\big/$$\bar{\mathsf{q}}_{\phi,t}\vhead{\ell}$ & Average input erasure prob. from $\uth{t}\head{i,\phi}$$\big/$$\vth{t}^\phi$ to $\fphi{t}$\\
			$\mathsf{p}_{i,\phi,t}\vhead{\ell}$$\big/$$\mathsf{q}_{\phi,t}\vhead{\ell}$ & Output extrinsic erasure prob. from $\fphi{t}$ to $\uth{t}\head{i,\phi}$$\big/$$\vth{t}^\phi$\\
			\bottomrule
			& \hfill \vphantom{{\large A }} {$i \in\{1,2\}$, $\phi\in\{\U,\L\}$}\\
		\end{tabular}
	\end{table}
	
	\subsection{Window Decoding Threshold Loss}
	Type-1 BCCs exhibit threshold loss under window decoding, even with a large window size $w$. As shown in \cite[Table 1]{zhu-bcc_window}, the full/window decoding threshold for the rate-1/3 type-1 BCC$(m=1)$ on the BEC is 0.6609/0.6553$^{(w \geq 4)}$. Such loss still exists for a large coupling memory such as $m=5$, where the full/window decoding threshold is 0.6650/0.6619$^{(w \geq 13)}$.
	
	\begin{figure}[t]
		\centering
		\subcaptionbox{ Type-1 BCC with $\mathbf{G}_{457}$ \label{fig:DE_bcc1_457_extr_evolution} }  {\includegraphics[width=8cm]{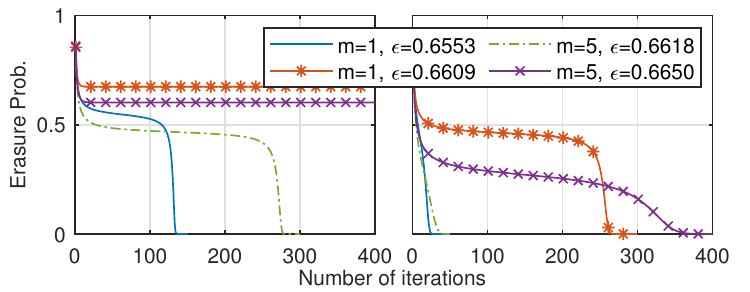}}
		\subcaptionbox{ Type-1 BCC with $\mathbf{G}_{537}$ \label{fig:DE_bcc1_537_extr_evolution} }  {\includegraphics[width=8cm]{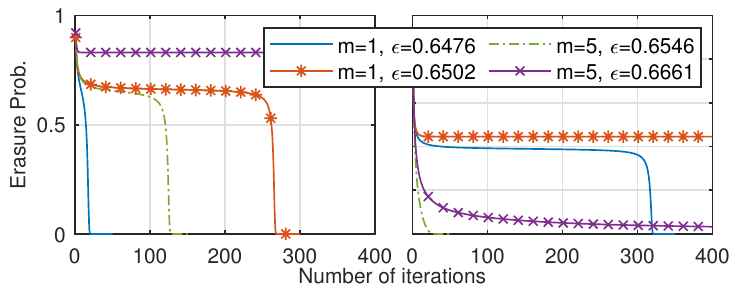}} 
		\subcaptionbox{ Type-2 BCC with  $\mathbf{G}_{457}$ \label{fig:DE_bcc2_457_extr_evolution} }  {\includegraphics[width=8cm]{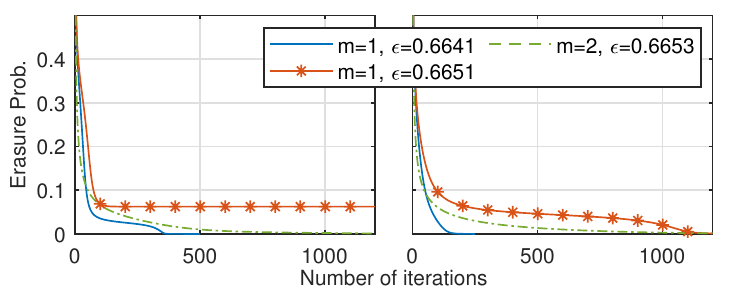}}
		\caption{Evolution of $\mathsf{p}_{1,\U,1}\vhead{\ell}$ (left subplots) and $\mathsf{p}_{1,\U,T}\vhead{\ell}$ (right subplots) for BCCs under full decoding.} 	\label{fig:DE_bcc1}
	\end{figure}
	
	For type-1 BCCs, the average input erasure probabilities $(\bar{\mathsf{p}}_{1,\U,t}\vhead{\ell}, \bar{\mathsf{p}}_{2,\U,t}\vhead{\ell}, \bar{\mathsf{q}}_{\U,t}\vhead{\ell})$ and $(\bar{\mathsf{p}}_{1,\L,t}\vhead{\ell}, \bar{\mathsf{p}}_{2,\L,t}\vhead{\ell}, \bar{\mathsf{q}}_{\L,t}\vhead{\ell})$ are derived in \cite[Sec. VD-2]{Moloudi-scTurbo}. The a-posterior erasure probability of the $t$-th source information sequence is $\mathsf{p}_{\ut{t}}\vhead{\ell} =\varepsilon \mathsf{p}_{1,\U,t}\vhead{\ell}  \mathsf{p}_{1,\L,t}\vhead{\ell} $.
	
	By tracking the evolution of the erasure probabilities, we observe that the threshold loss is because the amount of reliable decoding messages passed in the forward and backward directions could be unequal, e.g. the path $\vth{t-1}^\L \rightarrow \fu{t} \rightarrow \uth{t}\head{2,\U} $ and its reversal $\vth{t-1}^\L \leftarrow \fu{t} \leftarrow \uth{t}\head{2,\U}$ could have different message propagation capabilities.
	This implies that the maximum $\varepsilon$ required  for convergence at the head-side and tail-side of the coupling chain may be distinct.
	Knowing that DE starts with initial conditions: $\mathsf{p}_{i,\phi,t}\vhead{0}=0$ and $\mathsf{q}_{\phi,t}\vhead{0}=0$ for any $t \leq 0$ or $\geq T+1$. Let $\eh$/$\et$ be the head-/tail-side convergence threshold such that $\mathsf{p}_{\ut{1}}\vhead{\infty}$ approaches 0 iff $\varepsilon \leq \eh$, and $\mathsf{p}_{\ut{T}}\vhead{\infty}$ approaches 0 iff $\varepsilon \leq \et$.
	When a full decoder is used, the entire coupling chain can be successfully decoded as long as either side starts to converge, i.e. $\ebpf=\max(\eh, \et)$. When using a window decoder, the convergence  behavior of the first block determines the decoding result, so that $\ebpw=\eh$. Therefore, the threshold loss occurs  when $\eh<\et$.
	
	To illustrate the unbalanced convergence behavior, we plot  the erasure probability evolution of type-1 BCC with the component generator $ \Ga = \left[ 1, 0, \frac{1}{1+D+D^2}  ; 0, 1, \frac{1+D^2}{1+D+D^2} \right] $, which is commonly used in existing BCC papers \cite{Zhang-bcc,zhu-bcc_window,Moloudi-scTurbo} due to the optimized distance growth \cite{Zhang-bcc}.
	In particular, we compare the information extrinsic erasure probability against the number of iterations $\ell$ for the head-/tail-side of the BCC coupling chain, i.e. $\mathsf{p}_{1,\U,1}\vhead{\ell}$ and $\mathsf{p}_{1,\U,T}\vhead{\ell}$. At $\varepsilon=\et=0.6609$, the tail-side starts to converge. When $\varepsilon=\eh=0.6553$, the head-side converges while the tail-side shows a faster convergence.
	
	The convergence behavior is jointly influenced  by the choice of the component code generator $\mathbf{G}_\mathcal{C}$ and the coupling structure. To show the effect of $\mathbf{G}_\mathcal{C}$, Fig. \ref{fig:DE_bcc1_537_extr_evolution} plots the erasure probability evolution of type-1 BCCs with component generator $ \Gb = \left[ 1,  0, \frac{1+D^2}{1+D+D^2} ;   0,  1, \frac{D+D^2}{1+D+D^2} \right] $.
	We choose $\Gb$ because it has a good information weight property, and the uncoupled ensembles of BCCs with $\Gb$ also have good maximum a-posteriori (MAP) thresholds for a wide range of code rates (see Table \ref{tb:all_DE}). When using $\Gb$, the most significant change compared to $\Ga$ is that the $\eh$ becomes higher than $\et$ for $m=1$. To show the effect of coupling structure, Fig. \ref{fig:DE_bcc2_457_extr_evolution} plots the erasure probability evolution of type-2 BCC with $\Ga$, where $\eh$ increases to 0.6641 due to the additional decoding path brought by coupling information bits, even though $\et=0.6651$ is still higher. When $m=2$, both head-side and tail-side can converge at $\varepsilon=0.6653$, and the convergence speeds are nearly the same.
	
	To solve the window decoding threshold loss of BCCs, we need to enlarge the message propagation capability in the forward direction. Intuitively, this is done by two approaches: i) to choose a generator that allows the resultant type-1 BCC$(m=1)$ to achieve $\eh>\et$; and ii) to design a coupling structure that propagates more decoding messages via information coupling. 
	As shown in Sec. III, half coupling structure eliminates the randomness in coupling, as well as reduces the dependency between factor nodes and forms long decoding cycles to facilitate efficient message propagation. 
	
	\subsection{Density Evolution Analysis for Half SC-TCs} 
	\subsubsection{HSC-PCCs} \label{sec:de_hscpcc}
	The input erasure probabilities for single-sided SC-PCC$(m \geq 1)$ are expressed as follows:				
	\begin{subequations}\label{DE:HSC_PCC}
		\begin{align}
			& \textstyle \bar{\mathsf{p}}_{1,\U,t}\vhead{\ell}=\varepsilon  \Big(  \lambda_0 \mathsf{p}_{1,\L,t}\vhead{\ell-1}  +  \sum_{j=1}^m \lambda_j \big( \mathsf{p}_{1,\L,t+j}\vhead{\ell-1} \big) \Big) , \\
			& \textstyle \bar{\mathsf{p}}_{1,\L,t}\vhead{\ell}=\varepsilon\Big(  \lambda_0 \mathsf{p}_{1,\U,t}\vhead{\ell-1}  + \sum_{j=1}^m \lambda_j \big( \mathsf{p}_{1,\U,t-j}\vhead{\ell-1} \big) \Big) , \\
			& \bar{\mathsf{q}}_{\U,t}\vhead{\ell}= \bar{\mathsf{q}}_{\L,t}\vhead{\ell}= \ev .
		\end{align}
	\end{subequations}
	As proved in Appx. \ref{appx:proof}, the full BP decoding threshold of HSC-PCC can reach the MAP threshold when $\lambda_0=\hdots=\lambda_m=\frac{1}{m+1}$. DE computations also confirm that HSC-PCCs$(m=1)$ has $\ebp=\emap$ (see Table \ref{tb:scpcc} in Appx. \ref{appx:proof}).
	
	\subsubsection{HSC-BCCs}
	The input erasure probabilities for HSC-BCC$(\delta \geq 2)$ are expressed as
	\begin{subequations}
		\begin{align}
			& \textstyle \bar{\mathsf{p}}_{1,\tau}\vhead{\ell} = \frac{\varepsilon}{2} (\mathsf{p}_{1,\tau-\delta+1}\vhead{\ell-1} + \mathsf{p}_{1,\tau+\delta-1}\vhead{\ell-1}), \\
			& \bar{\mathsf{p}}_{2,\tau}\vhead{\ell} = \ev \mathsf{q}_{\tau-\delta }\vhead{\ell-1}  ,\\
			& \bar{\mathsf{q}}_{\tau}\vhead{\ell} = \ev \mathsf{p}_{2,\tau+\delta }\vhead{\ell-1} .
		\end{align}
	\end{subequations}
	
	Table \ref{tb:all_DE} shows comparisons of the decoding thresholds between different ensembles and decoders. The comparisons of $\eh$ and $\et$ can be found in Appx. B. For type-1 and type-2 BCCs, we compare $\ebpw$ given by different generators. Substituting $\Ga$ with other generators does not significantly improve $\ebpw$. The non-preferred SC-BCC variant in Fig. \ref{fig:factor_graph_hbcc1_notgood}, which straightly combines type-1 BCC and HSC-PCC, has a lower $\ebpw$ compared to type-2 BCC. HSC-BCC$(\delta=2)$ closely approaches $\emap$ of the uncoupled BCC and only shows only a slight loss at very high code rates. In addition, HSC-BCC ensembles also achieve better decoding thresholds under random puncturing and window decoding than SC-LDPC ensembles \cite{Mitchell-scldpc} without puncturing under full BP decoding by using a smaller coupling memory.
	
	\begin{table}[t]
		\centering
		\caption{Decoding thresholds for different code ensembles.} \label{tb:all_DE}
		\setlength{\tabcolsep}{2pt}
		\begin{tabular}{lccccccccc}
			\toprule
			\multicolumn{9}{c}{$\epsilon_{\text{BP,full}}$(SC-LDPC), $\emap$(UC-BCC), and $\ebpw$(SC-BCC)} \\ \cmidrule{1 - 9}
			Ensemble & $\boldsymbol{G}_{\mathcal{C}}$ & $m$ & \raisebox{2pt}{$_{R=1/3}$} & \raisebox{2pt}{$_{R=1/2}$}  & \raisebox{2pt}{$_{R=2/3}$}  & \raisebox{2pt}{$_{R=3/4}$}  & \raisebox{2pt}{$_{R=4/5}$}  & \raisebox{2pt}{$_{R=9/10}$} \\
			\midrule
			& $d_v=3$ & $2$ & - & 0.4881 & 0.3196 & 0.2373 & 0.1886 & 0.0930\\	
			{\scriptsize SC-LDPC} & $d_v=4$ & $3$ & 0.6656 & 0.4977 & 0.3302 & 0.2469 & 0.1971 & 0.0981\\	
			& $d_v=5$ & $4$ & - & 0.4994 & 0.3325 & 0.2491 & 0.1991 & 0.0994\\	
			\midrule
			& $\boldsymbol{G}_{457}$ & - & 0.6654 & 0.4992 & 0.3331 & 0.2499 & 0.1999 & 0.0999 \\
			%{\scriptsize BCC} & $\boldsymbol{G}_{547}$ & - &0.6654 & 0.4985 & 0.3326 & 0.2496 & 0.1998 & 0.0999 \\
			{\scriptsize UC-BCC} & $\boldsymbol{G}_{357}$ & - &0.6661 & 0.4996 & 0.3332 & 0.2499 & 0.1999 & 0.0999 \\
			& $\boldsymbol{G}_{537}$ & - & 0.6661 & 0.4993 & 0.3329 & 0.2498 & 0.1999 & 0.0999 \\
			\midrule
			{\scriptsize HSC-BCC} & $\boldsymbol{G}_{537}$ & $1$\raisebox{2pt}{$_{(\delta=2)}$} & 0.6661 & 0.4993 & 0.3329 & 0.2497 & 0.1997 & 0.0990 \\ 
			\cmidrule{2 - 9}
			& $\boldsymbol{G}_{357}$ & $1$\raisebox{2pt}{$_{(\delta=2)}$} & 0.6661 & 0.4996 & 0.3331 & 0.2498 & 0.1998 & 0.0990 \\
			&  &$2$\raisebox{2pt}{$_{(\delta=3)}$} & 0.6661 & 0.4996  & 0.3332 & 0.2499 & 0.1999 & 0.0996 \\				
			&  &$2$\raisebox{2pt}{$_{(\delta=4)}$} & 0.6661 & 0.4996  & 0.3332 & 0.2499 & 0.1999 & 0.0997 \\	
			\midrule
			{\scriptsize Type-1} & $\boldsymbol{G}_{457}$ & $1$ &0.6553& 0.4859& 0.3173 & 0.2408 & 0.1830 & 0.0843 \\	
			{\scriptsize BCC} & $\boldsymbol{G}_{537}$   & 			& 0.6502 & 0.4830 &  0.3162  & 0.2395 & 0.1833 & 0.0853\\	
			%				 					& $\boldsymbol{G}_{357}$   & 			& 0.6555& 0.4855 & 0.3169 & 0.2406 & 0.1828 & 0.0843\\	
			\midrule
			{\scriptsize Type-2} & $\boldsymbol{G}_{457}$ & $1$ & 0.6641 & 0.4973 & 0.3300 & 0.2455 & 0.1944 & 0.0914\\	
			{\scriptsize BCC} & $\boldsymbol{G}_{537}$   & 			& 0.6638 & 0.4957 & 0.3278 & 0.2435& 0.1928& 0.0911\\	
			\midrule
			{\scriptsize Fig. \ref{fig:factor_graph_hbcc1_notgood}} & $\boldsymbol{G}_{457}$ & $1$ & 0.6610 & 0.4925 & 0.3240 & 0.2402 & 0.1902 & 0.0911\\	
			\bottomrule
			\multicolumn{9}{r}{$^\star$Parity bits are punctured for BCCs with rates above 1/3}	 \vphantom{{\large A }}\\ 
		\end{tabular}
	\end{table}
	
	\section{Simulation Results} 
	Simulations are conducted over AWGN channels to verify the finite-length performance of HSC-BCCs (see Appx. \ref{appx:scpccber} for results of HSC-PCCs). All simulations on SC-TCs use linear-log-MAP algorithm with single-precision computation for RSC code decoding. 
	
	For type-1/type-2 BCCs, the generator matrix $\Ga$ is always used. Each horizontal iteration is a forward-backward round-trip, and the number of vertical iterations is fixed to 1. The decoding latency is $\mathcal{L}=4wI_hK$, where $I_h$ is the maximum number of horizontal iterations. 
	
\begin{figure}[H]
		\centering
		\includegraphics[width=7cm]{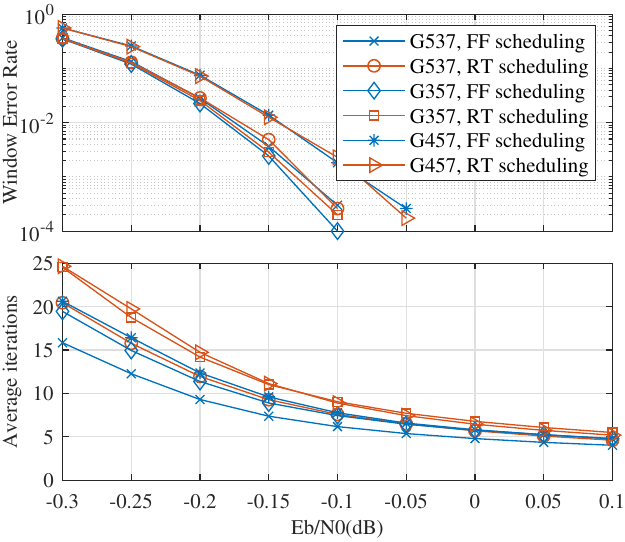}
		\caption{Convergence behavior of the first decoding window with $(K=8000,R=1/3,w=4)$ over AWGN channel.} 
		\label{fig:WER_ADI_K8000}
		\vspace{10pt}
%			\end{figure}
%		\begin{figure}[h]
%		\centering
		\subcaptionbox{$K=8000$ and $R=1/3$. \label{fig:BER_R13_K8000} }  {\includegraphics[width=7cm]{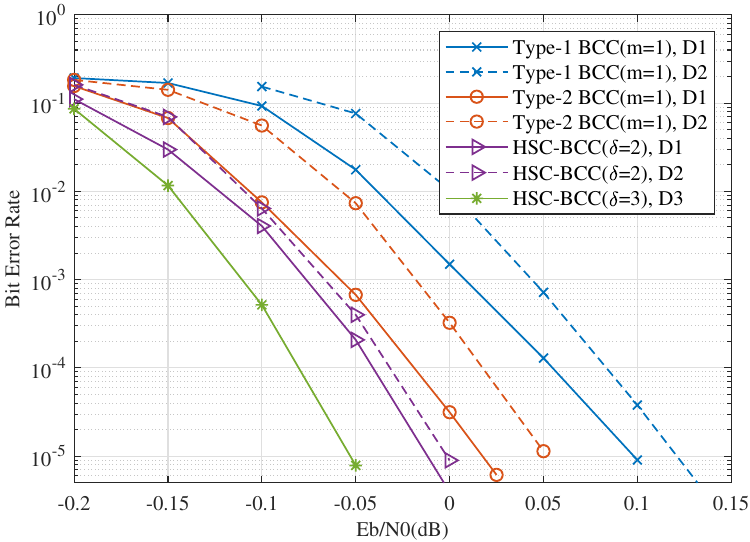}} 
		\subcaptionbox{$K=6000$ and $R=1/2$. \label{fig:BER_R12_K6000} }  {\includegraphics[width=7cm]{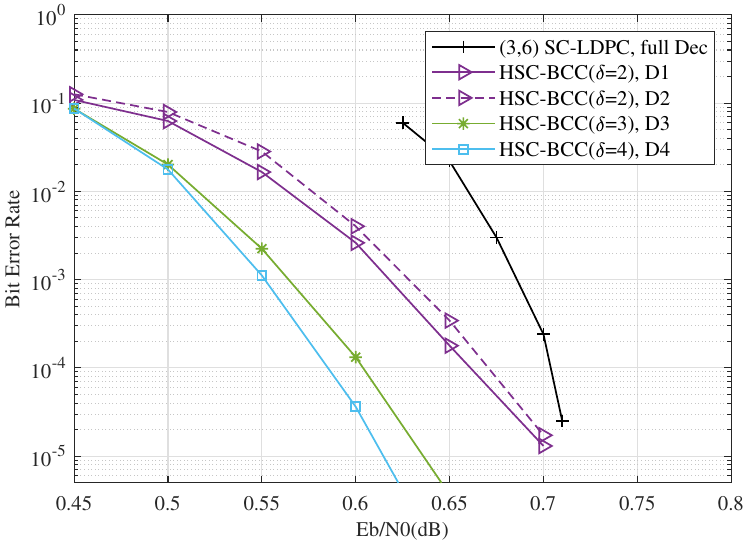}}
		\caption{Error performance of BCCs over AWGN channel. HSC-BCCs use $\Gb$, and type-1/type-2 BCCs use $\Ga$. $T=50$. Decoding parameter${(w,I_h)}$: D1-${(4,20)}$, D2-${(4,10)}$, and D3-${(6,13)}$, and D4-${(8,10)}$. SC-LDPC BER curve is from \cite[Fig. 19]{Mitchell-scldpc}. } \label{fig:BER}
	\end{figure}

	First, we investigate the window decoding scheme of HSC-BCCs. In Fig. \ref{fig:WER_ADI_K8000}, we compare the convergence behavior of the first decoding window, which outputs the hard-decision estimation of the first $K=8000$ information bits $[\hat{\bm{u}}_{\tau=1}^\prime,\hat{\bm{u}}_{\tau=2}^\prime]$. The upper subplot depicts the error rate of the first window  $\text{WDER} \triangleq \Pr\{[\hat{\bm{u}}_{\tau=1}^\prime,\hat{\bm{u}}_{\tau=2}^\prime] \neq [\utp{\tau=1},\utp{\tau=2}] \}$ after $I_h=100$ iterations. The lower subplot depicts the average number of iterations needed for convergence per successful decoding attempt. 
	Two decoding schedules are compared: forward-backward round-trip scheduling (marked as ``RT scheduling''), and forward decoding twice per iteration (marked as ``FF scheduling''). One can see that $\Gb$ and $\Gbf$ give better WDER than $\Ga$ as expected by the DE analysis. With the same generator, although the two schedulings have similar WDER, FF scheduling has fewer iterations for successful decoding attempts. This suggests that FF scheduling would bring faster convergence speed, which is valuable for reducing decoding latency. This observation aligns with our design target, which is to increase the amount of reliable decoding messages passed in the forward direction. Besides, although $\Gbf$ has better thresholds over $\Gb$, it shows a slower convergence speed. Thus, we choose $\Gb$ to build HSC-BCCs in the other simulation results.

	Fig. \ref{fig:BER} investigates the BER performance of HSC-BCCs at finite blocklengths.  Coupling length ${T=50}$ is used. Fig. \ref{fig:BER}a compares the BER of type-1/type-2 BCC$({m=1})$ with decoder D1, HSC-BCC${(\delta=2)}$ with decoder D1, and HSC-BCC${(\delta=3)}$ with decoder D3. At similar latency levels, both HSC-BCCs outperform the other two BCCs. In addition, when $I_h$ drops from $20$ to $10$, HSC-BCC${(\delta=2)}$ has less BER loss than type-1 and type2-BCCs. This implies that HSC-BCCs have a faster convergence speed. Fig. \ref{fig:BER}b shows that HSC-BCCs under window decoding outperform the $(3,6)$-regular photograph SC-LDPC code \cite{Mitchell-scldpc} under full decoding of the whole coupled chain. In addition, although increasing the delay factor $\delta$ does not bring a very significant decoding threshold gain at both rates 1/3 and 1/2, the BER performance improves with $\delta$.

	\section{Conclusions}  
	This paper introduced HSC-PCCs and HSC-BCCs to simplify SC-PCCs and type-2 BCCs in \cite{Moloudi-scTurbo}. With a properly designed deterministic coupling structure, both codes can be constructed using encoders consisting of fewer interleaving and (de-)multiplexing components than SC-PCCs and type-2 BCCs. With such a simple structure, both codes are close to the MAP decoding threshold of their uncoupled ensemble when having small coupling memory. Moreover, HSC-BCCs were carefully designed to approach full decoding thresholds under window decoding. At finite blocklengths and under window decoding, HSC-BCCs demonstrated better error performance than type-1/type-2 BCCs while having faster convergence. 
	
	\clearpage
	%	\bibliographystyle{ieeetr}
	%	\bibliography{hscbcc_reference}

	\clearpage
	
	\begin{appendices}
		\section{Proof of Threshold Saturation for single-sided SC-PCCs} \label{appx:proof}
		For the single-sided SC-PCC ensembles with $\lambda_0=\hdots=\lambda_m=\frac{1}{m+1}$, i.e. $\lambda = \frac{m}{m+1}$ and coupled subsequences are of equal lengths, the DE equations for information bits in \eqref{DE:HSC_PCC} become
		\begin{subequations}\label{DE:HSC_PCC1}
			\begin{align}
				& \bar{\mathsf{p}}_{1,\U,t}\vhead{\ell}=F_1\Bigg( \frac{\varepsilon}{m+1}\sum^m_{j=0}  \mathsf{p}_{1,\L,t+j}\vhead{\ell-1}  , \ev\Bigg), \\
				& \bar{\mathsf{p}}_{1,\L,t}\vhead{\ell}=F_1\Bigg(\frac{\varepsilon}{m+1}\sum^m_{j=0}  \mathsf{p}_{1,\U,t-j}\vhead{\ell-1} ,\ev\Bigg).
			\end{align}
		\end{subequations}
		Note that unlike the conventional SC-PCCs \cite{Moloudi-scTurbo}, the coupling structures in the upper and lower RSC encoders are asymmetric. Let $x_t\vhead{\ell} \triangleq \frac{1}{m+1}\sum^m_{j=0}  \mathsf{p}_{1,\U,t-j}\vhead{\ell}$. We can rewrite the DE equations \eqref{DE:HSC_PCC1} into
		\begin{subequations}
			\begin{align}
				x_t\vhead{\ell}=&\frac{1}{m+1}\sum^m_{j=0}F_1\Bigg(\frac{\varepsilon}{m+1}\sum^m_{k=0}F_1\Big(\varepsilon x_{t-j+k}\vhead{\ell-2} ,\ev\Big),\ev\Bigg) \\
				=&\frac{1}{m+1}\sum^m_{j=0}f\Bigg(\frac{1}{m+1}\sum^m_{k=0}g\Big(x_{t-j+k}\vhead{\ell-2} \Big); \varepsilon\Bigg),\label{eq:trs_pcc1}
			\end{align}
		\end{subequations}
		where \eqref{eq:trs_pcc1} follows by letting $f(x;\varepsilon) =g(x)= F_1(\varepsilon x,\ev)$. It is easy to verify that $f(x;\varepsilon)$ and $g(x)$ form a scalar admissible system \cite[Def. 1]{Yedla-threshold_satruation}. By \cite[Th. 1]{Yedla-threshold_satruation}, for any $\varepsilon < \varepsilon_{\text{MAP}}$, the only fixed point of the recursion in \eqref{eq:trs_pcc1} is $\boldsymbol{x}\vhead{\infty} = \big[ x_t\vhead{\infty}, \hdots ,x_T\vhead{\infty} \big] = \boldsymbol{0}$ as $T \rightarrow \infty$, $m \rightarrow \infty$, and $T \gg m$.
		
		Table \ref{tb:scpcc} compares the BP decoding thresholds $\ebp$ of single-sided SC-PCC$(m=1)$ with the MAP decoding thresholds $\emap$ of the uncoupled ensemble. One can see that the BP decoding threshold of single-sided SC-PCC$(m=1,\lambda_1=0.5)$, i.e. the HSC-PCC$(m=1)$, approaches the MAP decoding thresholds for all considered rates. 
		
		\begin{table}[H]
			\caption{$\emap$ of the uncoupled ensemble of single-sided SC-PCCs, and $\ebp$ of single-sided SC-PCCs for different $\lambda_1$.} \label{tb:scpcc}
			\begin{tabular}{lllllll}
				\toprule
				\multicolumn{1}{c}{ } & \multicolumn{6}{c}{$\emap$ or $\ebp\head{\lambda_1}$ at rate $R$} \\ \cmidrule{2 - 7}
				& \raisebox{2pt}{$_{R=1/3}$} & \raisebox{2pt}{$_{R=1/2}$}  & \raisebox{2pt}{$_{R=2/3}$}  & \raisebox{2pt}{$_{R=3/4}$}  & \raisebox{2pt}{$_{R=4/5}$}  & \raisebox{2pt}{$_{R=9/10}$} \\ 
				\midrule 
				$\emap$ & 0.6621 & 0.4863 & 0.3080 & 0.2204 & 0.1698 & 0.0769 \\
				\midrule 
				$\lambda_1=0$ 	& 0.6368 & 0.4651 & 0.2945 & 0.2111 & 0.1628 & 0.0740 \\
				$\lambda_1=0.1$ & 0.6462 & 0.4743 & 0.3016 & 0.2164 & 0.1671 & 0.0760 \\
				$\lambda_1=0.2$ & 0.6533 & 0.4805 & 0.3054 & 0.2191 & 0.1690 & 0.0766 \\
				$\lambda_1=0.3$ & 0.6579 & 0.4839 & 0.3071 & 0.2200 & 0.1696 & 0.0768 \\
				$\lambda_1=0.4$ & 0.6604 & 0.4855 & 0.3077 & 0.2203 & 0.1697 & 0.0768 \\
				$\lambda_1=0.5$ & 0.6621 & 0.4863 & 0.3080 & 0.2204 & 0.1698 & 0.0769 \\
				\bottomrule 
			\end{tabular}
		\end{table}
		
		\newpage
		
		\section{Convergence thresholds $\eh$ and $\et$} \label{appx:eh_et}
		Table \ref{tb:Eh_Et} compares the convergence thresholds for BCCs. Noteworthy, HSC-BCCs always have $\eh>\et$ regardless of generator $\boldsymbol{G}_{\mathcal{C}}$, aligning with our design target. 
		
		\begin{table}[H]
			\centering
			\caption{Head and tail convergence thresholds $(\eh,\et)$ under full decoding}
			\label{tb:Eh_Et}
			\setlength{\tabcolsep}{2pt}
			\begin{tabular}{lccccccccc}
				\toprule
				\multicolumn{1}{c}{ } &  &   & \multicolumn{6}{c}{ $\eh$ and $\et$ at rate $R$} \\ \cmidrule{4 - 9}
				Ensemble & $\boldsymbol{G}_{\mathcal{C}}$ & $\epsilon$ & \raisebox{2pt}{$_{R=1/3}$} & \raisebox{2pt}{$_{R=1/2}$}  & \raisebox{2pt}{$_{R=2/3}$}  & \raisebox{2pt}{$_{R=3/4}$}  & \raisebox{2pt}{$_{R=4/5}$}  & \raisebox{2pt}{$_{R=9/10}$} \\
				\midrule			
				% --------------------------------------
				{\scriptsize HSC-BCC} & $\boldsymbol{G}_{457}$ & $\eh$ & 0.6653 & 0.4992 & 0.3331 & 0.2498 & 0.1998 & 0.0990 \\
				${(\delta=2)}$ & & $\et$ & 0.6609 & 0.4932 & 0.3257 & 0.2419 & 0.1915 & 0.0902 \\
				\cmidrule{2 - 9} % --------------
				& $\boldsymbol{G}_{537}$ & $\eh$ & 0.6661 & 0.4993 & 0.3329 & 0.2497 & 0.1997 & 0.0990 \\ 
				& & $\et$ & 0.6476 & 0.4770 & 0.3085 & 0.2256 & 0.1765 & 0.0816 \\
				\cmidrule{2 - 9}% --------------
				& $\boldsymbol{G}_{357}$ & $\eh$ & 0.6661 & 0.4996 & 0.3331 & 0.2498 & 0.1998 & 0.0990 \\
				& & $\et$ & 0.6432 & 0.4696 & 0.3008 & 0.2195 & 0.1721 & 0.0815 \\  
				% --------------------------------------
				\midrule
				{\scriptsize Type-1} & $\boldsymbol{G}_{457}$ & $\eh$ &0.6553& 0.4859& 0.3173 & 0.2408 & 0.1830 & 0.0843 \\	
				{\scriptsize BCC} & & $\et$ & 0.6609 & 0.4932 & 0.3257 & 0.2419 & 0.1915 & 0.0902\\
				\cmidrule{2 - 9} % --------------
				${(m=1)}$ & $\boldsymbol{G}_{537}$   & $\eh$ & 0.6502 & 0.4830 &  0.3162  & 0.2395 & 0.1833 & 0.0853\\	
				& & $\et$ & 0.6476 & 0.4770 & 0.3085 & 0.2256 & 0.1765 & 0.0816\\ 
				\cmidrule{2 - 9} % --------------
				& $\boldsymbol{G}_{357}$ & $\eh$ & 0.6555 & 0.4855 & 0.3169 & 0.2330 & 0.1828 & 0.0843 \\
				& & $\et$ & 0.6432 & 0.4696 & 0.3008 & 0.2195 & 0.1721 & 0.0815 \\
				\midrule
				% --------------------------------------
				{\scriptsize Type-2} & $\boldsymbol{G}_{457}$ & $\eh$ & 0.6641 & 0.4973 & 0.3300 & 0.2455 & 0.1944 & 0.0914\\	
				{\scriptsize BCC} & & $\et$ & 0.6651 & 0.4988 & 0.3323 & 0.2488 & 0.1986 & 0.0964 \\
				\cmidrule{2 - 9}
				${(m=1)}$ & $\boldsymbol{G}_{537}$   & $\eh$ & 0.6638 & 0.4957 & 0.3278 & 0.2435& 0.1928& 0.0911\\	
				& & $\et$ & 0.6561 & 0.4870 & 0.3194 & 0.2361 & 0.1864 & 0.0880 \\
				\cmidrule{2 - 9} % --------------
				& $\boldsymbol{G}_{357}$ & $\eh$ & 0.6645 & 0.4973 & 0.3298 & 0.2454 & 0.1943 & 0.0914 \\ 
				& & $\et$ & 0.6570 & 0.4861 & 0.3158 & 0.2318 & 0.1823 & 0.0869 \\ 
				\bottomrule
				\multicolumn{9}{r}{$^\star$Parity bits are punctured for rates above 1/3}	 \vphantom{{\large A }}\\ 
			\end{tabular}
			\label{tb:dh_dt}
		\end{table}
		
		\section{HSC-PCC Simulation results} \label{appx:scpccber}
\begin{figure}[h]
			\centering
			\includegraphics[width=7cm]{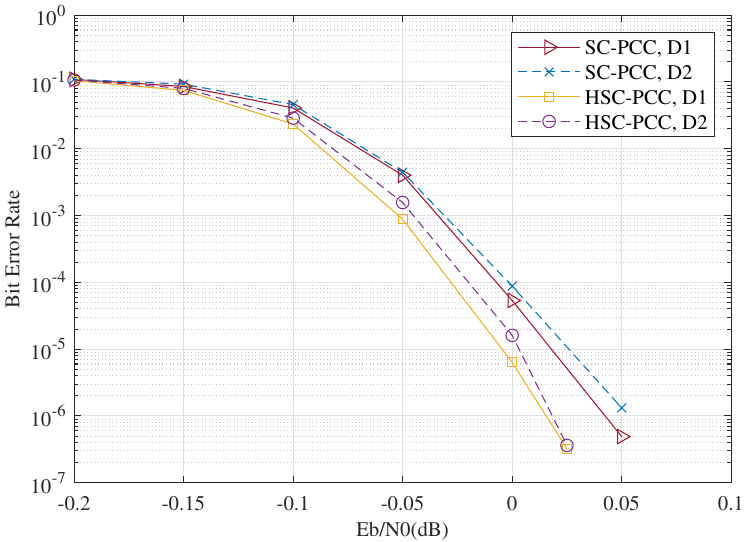}
			\caption{Error performance of SC-PCCs and HSC-PCCs over AWGN channel. Generator $\mathbf{G}=[1, 15/13]$, $R=1/3$, $T=50$, $K=8000$. SC-PCC$(m=1)$ uses the original encoder proposed in \cite{Moloudi-scTurbo}. Decoding parameter${(w,I_h)}$: D1-${(8,20)}$ and D2-${(8,10)}$. }
			\label{fig:BER_hscpcc}
		\end{figure}

		Fig. \ref{fig:BER_hscpcc} compares the error performance of SC-PCC$(m=1)$ and HSC-PCC$(m=1)$ at $R=1/3$. With a simplified substructure, HSC-PCCs have slightly better error performance compared to SC-PCCs. 
		
	\end{appendices}
	
\end{document}